\documentclass[twocolumn, showpacs,preprintnumbers,amsmath,amssymb]{revtex4}

\usepackage{graphicx}

\begin{document}

\title{Entanglement in finite spin rings with noncollinear Ising interaction}

\author{F. Troiani}

\affiliation{Istituto Nanoscienze-CNR, S3, Modena, Italy}

\date{\today}

\begin{abstract}

We investigate the entanglement properties of finite spin rings, with 
noncollinear Ising interaction between nearest neighbours. 
The orientations of the Ising axes are determined either by the spin 
position within the ring (model $A$) or by the direction of the bond 
(model $B$). In both cases, the considered spin Hamiltonians have a
point group symmetry, rather than a translation invariance, as in spin 
rings with collinear Ising interaction. The ground state of these 
models exhibit remarkable entanglement properties, resembling  
GHZ-like states in the absence of an applied magnetic field (model $B$). 
Besides, the application of an homogeneous magnetic field allows to 
modify qualitatively the character of the ground state entanglement, 
switching from multipartite to pairwise quantum correlations 
(both models $A$ and $B$).

\end{abstract}


\maketitle

\section{Introduction}

Spin rings represent prototypical low-dimensional systems with highly entangled
ground states \cite{RevModPhys.80.517,RevModPhys.81.865}. In particular,
antiferromagnetic and isotropic exchange interaction induces pairwise entanglement 
between nearest-neighbouring spins; moreover, it maximizes the concurrence 
within the set of translationally invariant states with vanishing magnetization
\cite{PhysRevA.63.052302}. 
Quantum entanglement in anisotropic Heisenberg models has also been investigated,
partly in relation to the separability of the ground state for specific values of
the applied magnetic field \cite{Kurmann1982235,PhysRevLett.100.197201,
PhysRevB.81.054415,PhysRevA.77.052322,PhysRevA.74.022322,PhysRevA.77.012319,
PhysRevLett.93.167203}.
Indeed, the magnetic field can be used as a control parameter in order to 
engineer the ground state and thermal entanglement of the system. 
The interplay between the system anisotropies and the field offer even wider
possibilities if one assumes that this needs not be homogeneous, but can 
rather be controlled locally \cite{PhysRevA.80.062325,PhysRevB.81.054415,
PhysRevLett.100.100502}.

\begin{figure}[ptb]
\begin{center}
\includegraphics[width=8.5cm]{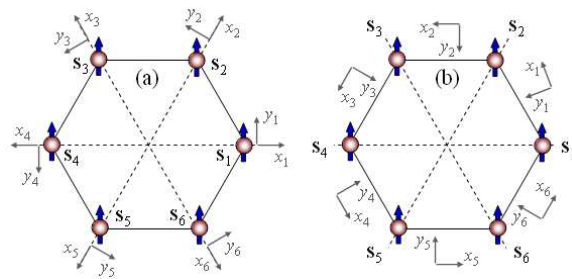}
\end{center}
\caption{(color online) Twisted-spin representations of an hexagonal spin ring.
(a) In the case of model $A$, the components of each spin $ {\bf s}_k $ in the 
Hamiltonian ($H^A$, Eq. \ref{eqha}) refer to local reference frames, with the 
$x_k$ axis pointing 
in the radial direction. The general reference frame is chosen such that 
$ \hat{\bf r}_1 = \hat{\bf x} $, being ${\bf r}_k$ the position of ${\bf s}_k$.
(b) In the case of model $B$, the local reference frames refer to each of the 
couplings between nearest neighbouring spins; the $x$ and $y$
components are thus defined for each bond, with $ \hat{\bf x}_k \parallel 
({\bf r}_{k+1} - {\bf r}_{k}) $. Therefore, the components of each $ {\bf s}_k $ in 
the Hamiltonian ($H^B$, Eq. \ref{eqhb}) refer to two local reference frames, 
one for the coupling with $ {\bf s}_{k-1} $ and one for that with 
$ {\bf s}_{k+1} $.}
\label{fig01}
\end{figure}
Previous analyses were mainly devoted to systems with translational invariance, 
where the anisotropies in the spin-spin couplings are independent on the 
site. 
However, physical implementations of low-dimensional spin systems are  
typically characterized by point-group symmetries, rather than traslational 
invariance \cite{tsukerblat}.
This is the case, for examples, of nanomagnets \cite{gatteschi}, that 
represent a rich class of molecular spin clusters, 
with widely tunable geometries and physical parameters. 
A number of ring-shaped nanomagnets has been investigated in the last 
years \cite{gatteschi};
some of these possesses attractive features for the encoding and manipulation 
of quantum information \cite{PhysRevLett.94.207208,PhysRevLett.94.190501,PhysRevLett.101.217201}. 
In this paper we consider rings formed by equivalent spins, where 
the anisotropies in the spin-spin couplings reflect the point-group 
symmetry of the molecule and the local environment of each spin. 
In particular, we focus on spin models with noncollinear Ising interaction 
between nearest neighbours \cite{PhysRevLett.100.247205,JACS2003}. 
The resulting spin Hamiltonians don't fall into any of the commonly 
considered cases and - unlike the standard Ising model - can exhibit 
highly entangled ground states, also in the absence of an applied magnetic 
field. 

\section{The model} 

In order to provide an intuitive picture of the spin ring symmetry, 
we introduce the anisotropic Heisenberg model in a twisted-spin
representation. 
In particular, we consider the case where the spins ${\bf s}_i$ are 
located at the vertices of a regular polygon, and the 
directions of the coordinate axes are determined either by the position (${\bf r}_i$) 
of each spin within the ring (model $A$) or by the direction of the bond 
between the exchange-coupled
spins (model $B$). In the case of model $A$ [Fig. \ref{fig01}(a)], the
$ x_i $ component of spin ${\bf s}_i$ is along the radial 
direction ($ \hat{\bf r}_i = \hat{\bf x}_i $, with the origin $O$ corresponding
to the center of the polygon), whereas the $ y_i $ axis is defined 
so as to form with $x_i$ a left-handed reference frame in the polygon plane. 
In the case of model $B$
[Fig. \ref{fig01}(b)], instead, the $x$ components of two neighbouring 
and coupled spins,
${\bf s}_i$ and ${\bf s}_{i+1}$, are both along the side of the polygon
[$ \hat{\bf x}_i \parallel ({\bf r}_{i+1}-{\bf r}_i) $].

%
\begin{figure}[ptb]
\begin{center}
\includegraphics[width=8.5cm]{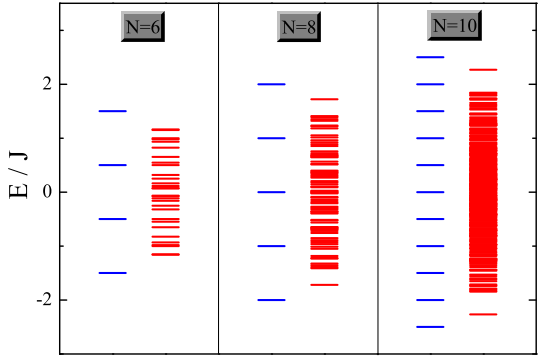}
\end{center}
\caption{(color online) Energy spectra of the noncollinear Ising Hamiltonians
$H^A_\xi $ (blue) and $H^B_\xi $ (red) with $N$ spins $s=1/2$ and $ \xi = X,Y $.  
For both model $A$ and model $B$, the spectrum is independent on whether 
$ J^\chi_{xx} = 0 \neq J^\chi_{yy} $ or $ J^\chi_{yy} = 0 \neq J^\chi_{xx} $
and on the sign of the coupling. While the ground states of $H^A_\xi $ are twofold
degenerate, the ground doublets of $H^B_\xi $ present the following splittings 
$ \delta $ (not appreciable in the figure): $ \delta / J \simeq 10^{-2}, 2.0 \times 
10^{-4}, 10^{-5} $, 
for $N=6,8,10$, respectively.}
\label{fig02}
\end{figure}

\subsection{Spin model A}

In model $A$, the anisotropies in the Heisenberg coupling reflect the position 
of each spin within the ring: 
\begin{eqnarray}\label{eqha}
H^A = \sum_{k=1}^N \left( J_{xx}^A s_{k,x}'\, s_{k+1,x}' +
                          J_{yy}^A s_{k,y}'\, s_{k+1,y}' \right) ,
\end{eqnarray}
where $N$ is the number of spins and $ {\bf s}_{N+1} \equiv {\bf s}_1 $. 
The primed spin components can be expressed in terms of the general 
reference frame: 
\begin{eqnarray}
s_{k,x}' & = & {\bf s}_k \cdot \hat{\bf x}_k^A = \cos \phi_k\, s_{k,x} + \sin \phi_k\, s_{k,y} ,\nonumber \\ 
s_{k,y}' & = & {\bf s}_k \cdot \hat{\bf y}_k^A = \cos \phi_k\, s_{k,y} - \sin \phi_k\, s_{k,x} ,
\end{eqnarray}
where 
$ \hat{\bf x}_k^A = ( \cos\phi_k , \sin\phi_k ) $
and
$ \hat{\bf y}_k^A = (-\sin\phi_k , \cos\phi_k ) $
in the $xy$ basis,
with $ \phi_k = 2(k-1)\pi / N $.
The Hamiltonian $ H^A $ can be thus rewritten in terms of the general reference
frame:
\begin{eqnarray}
H^A = \sum_{k=1}^N \sum_{\alpha , \beta = x , y} 
\left( J^{\alpha\beta}_k + D^{\alpha\beta}_k \right)\, s_{k,\alpha}\, s_{k+1,\beta} ,
\label{leq01}
\end{eqnarray}
where the tensors $ {\bf J} $ and $ {\bf D} $ account for the symmetric and 
antisymmetric components of the interaction ($ J^{\alpha\beta}_k \equiv J^{\beta\alpha}_k $
and 
$ D^{\alpha\beta}_k \equiv -D^{\beta\alpha}_k $), respectively:
\begin{eqnarray}
J^{xx}_k & = & J_{xx}^A \cos \phi_k\,\cos \phi_{k+1}\, + J_{yy}^A \sin \phi_k\,\sin \phi_{k+1}, 
\nonumber\\
J^{yy}_k & = & J_{yy}^A \cos \phi_k\,\cos \phi_{k+1}\, + J_{xx}^A \sin \phi_k\,\sin \phi_{k+1}, 
\nonumber\\
J^{xy}_k & = & \frac{1}{2} ( J_{xx}^A - J_{yy}^A) \sin (\phi_{k+1}+\phi_k), 
\nonumber\\
D^{xy}_k & = & \frac{1}{2} ( J_{xx}^A + J_{yy}^A) \sin (\phi_{k+1}-\phi_k),
\end{eqnarray}
while $ D_{xx} = D_{yy} = 0 $.

\begin{figure}[ptb]
\begin{center}
\includegraphics[width=8.5cm]{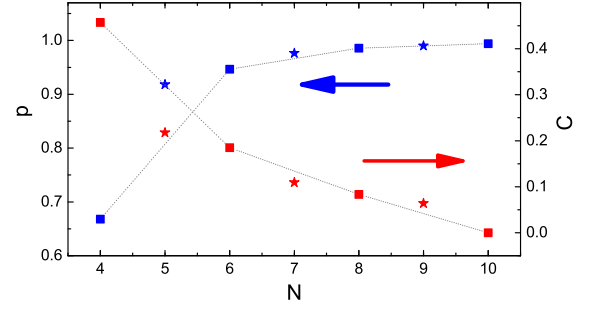}
\end{center}
\caption{(color online) Concurrence (red symbols, right axis) and squared modulus 
(blue symbols, left axis) of the overlap between the ground state 
of $H^B_\chi $ and the state $ | \widetilde{\Psi}^\xi (\phi) \rangle $ 
(with $\xi = F, AF$ and $\chi = X,Y$; see Eq. \ref{eqpsiGHZ}). For even spin number (squares), 
the values of $ p $ and $ C $ are independent 
on the direction and sign of the coupling, but different angles $ \phi $ are 
considered in the $\chi = Y $ ($ \phi = 0$) and $ \chi = X $ ($ \phi = \pi /2$) 
cases. In the case of odd spin numbers (stars) the $F$ cases have been 
considered. }
\label{fig03}
\end{figure}
In the case $ J_{xx}^A = J_{yy}^A \equiv J^A $, the couplings of the above 
Hamiltonian become site-independent, and they reduce to an isotropic symmetric 
contribution
($ J^{xx}_k = J^{yy}_k = J^A \cos (2\pi/N) $
plus an antisymmetric one
($ D^{xy}_k = J^A \sin (2\pi/N) $).
In the following, we shall focus on the noncollinear Ising models
$H^A_X$ and $H^A_Y$, 
corresponding to $J_{yy}^A=0\neq J_{xx}^A \equiv J^A$ and 
$J_{xx}^A=0\neq J_{yy}^A \equiv J^A$, respectively. Both $H^A_X$ and $H^A_Y$
include, in the general reference frame representation,
both (anisotropic) symmetric and antisymmetric terms, with components that 
depend on the spin index. 

\subsection{Spin model B}

In model $B$, the preferential directions in the anisotropic Heisenberg
model are determined by the direction of each bond:
\begin{eqnarray}\label{eqhb}
H^B = \sum_{k=1}^N \left( J_{xx}^B s_{k,x}'' s_{k+1,x}''' +
                          J_{yy}^B s_{k,y}'' s_{k+1,y}''' \right) .
\end{eqnarray}
Here, the components of each spin $ {\bf s}_k $ are referred to two local 
reference frames: in the couplings with the spins $ {\bf s}_{k-1} $ 
and $ {\bf s}_{k+1} $, these are $ (x_{k-1},y_{k-1}) $ and $ (x_{k},y_{k}) $, respectively:  
\begin{eqnarray}
s_{k  ,x}''  & = & {\bf s}_k \cdot \hat{\bf x}_k^B = 
\cos \varphi_k\, s_{k,x} + \sin \varphi_k\, s_{k,y} ,
\nonumber \\ 
s_{k  ,y}''  & = & {\bf s}_k \cdot \hat{\bf y}_k^B = 
\cos \varphi_k\, s_{k,y} - \sin \varphi_k\, s_{k,x} ,
\nonumber \\
s_{k+1,x}''' & = & {\bf s}_k \cdot \hat{\bf x}_k^B = 
\cos \varphi_k\, s_{k+1,x} + \sin \varphi_k\, s_{k+1,y} ,
\nonumber \\ 
s_{k+1,y}''' & = & {\bf s}_k \cdot \hat{\bf y}_k^B = 
\cos \varphi_k\, s_{k+1,y} - \sin \varphi_k\, s_{k+1,x} ,
\end{eqnarray}
being
$ \hat{\bf x}_k^B = ( \cos\varphi_k , \sin\varphi_k ) $,
$ \hat{\bf y}_k^B = (-\sin\varphi_k , \cos\varphi_k ) $
and
$ \varphi_k = \pi / 2 + (2k-1)\pi / N $.
In the general reference frame, the above Hamiltonian can be written as in 
Eq. \ref{leq01}, where:
\begin{eqnarray}
J^{xx}_k & = & J_{xx}^B \cos^2 \varphi_k\, + J_{yy}^B \sin^2 \varphi_k , 
\nonumber\\
J^{yy}_k & = & J_{yy}^B \cos^2 \varphi_k\, + J_{xx}^B \sin^2 \varphi_k , 
\nonumber\\
J^{xy}_k & = & (J_{xx}^B - J_{yy}^B) \cos \varphi_k\,\sin \varphi_k ,
\nonumber\\
D^{xy}_k & = & 0 .
\end{eqnarray}
In the case $ J_{xx}^B = J_{yy}^B \equiv J^B $, the couplings of $H^B$
become site-independent, and only the symmetric and isotropic contribution is retained:
$ J^{xx}_k = J^{yy}_k = J^B $, with $ J^{xy}_k = D^{xy}_k = 0 $.
In the following, we shall focus on the noncollinear Ising models
$H^B_X$ and $H^B_Y$, 
defined as $J_{yy}^B=0\neq J_{xx}^B \equiv J^B$ and 
$J_{xx}^B=0\neq J_{yy}^B \equiv J^B$, respectively.
These cases correspond to a symmetric 
exchange ($ D^{xy}_k = 0 $), with the principal directions that vary from one spin pair 
to another. 

We finally note that the above models $A$ and $B$ cannot, in general, be rephrased
one in terms of the other. In fact, model $A$ can be rewritten in the 
twisted-spin representation $B$ only by adding an antisymmetric contribution:
\begin{eqnarray}
J^B_{xx} & = & - J_{xx}^A \sin^2 (\pi / N) + J_{yy}^A \cos^2 (\pi / N) , 
\nonumber\\
J^B_{yy} & = &   J_{xx}^A \cos^2 (\pi / N) - J_{yy}^A \sin^2 (\pi / N) , 
\nonumber\\
J^B_{xy} & = & 0 ,
\nonumber\\
D^B_{xy} & = & (J_{xx}^A + J_{yy}^A) \cos (\pi / N) \sin (\pi / N) .
\end{eqnarray}
Analogous considerations apply to the model $B$ in the twisted-spin representation
$A$. In this case, the equations can be obtained from the above ones by swapping
the $A$ and $B$ apices, and by changing the sign in the expression of the 
anti-symmetric exchange coefficient.

\subsection{Symmetry properties of the $A$ and $B$ models}

Both the $A$ and $B$ models belong to the ${\bf D}_{nh}$ point-group 
symmetry \cite{tsukerblat},
with $n$ corresponding to the spin number ($ n = N $). 
In fact, one can show that $H^A$ and $H^B$ are invariant under $N$ different
$\hat{C}_2$ rotations $ \exp ( -i {\bf J} \cdot \hat{\bf n}_k \pi / \hbar ) $,
with $ {\bf J} = \sum_{i=1}^N ({\bf l}_i + {\bf s}_i) $: 
$N/2$ rotations have axes parallel to the spin positions, 
$ \hat{\bf n}_k = {\bf r}_k$ with $k=1,2,\dots , N/2$; 
the other $N/2$ rotations have axes that coincide with the bisectors of 
the polygon sides, 
$ \hat{\bf n}_k = ( {\bf r}_k + {\bf r}_{k+1} ) / | {\bf r}_k + {\bf r}_{k+1} | $, 
with $k=N/2+1,\dots , N$.
The Hamiltonians $H^A$ and $H^B$ are also invariant under reflection 
($\hat{\sigma}_h$) about the 
polygon plane $xy$, and under the
$ \hat{C}_n $ rotation 
$ \exp ( -i 2\pi J_z / N \hbar ) $
around the vertical axis $z$. 
We note that the collinear $XY$ model, 
$ H_{XY} = J_{xx} \sum_{k=1}^N ( s_{k,x'} s_{k+1,x'} + s_{k,y'} s_{k+1,y'}) $, 
belongs to the ${\bf D}_{2h}$ point-group symmetry, if the $x'$ and $y'$ axes 
coincide with symmetry axes of the polygon defined by the spin positions, 
i.e. if $ \hat{\bf x}'$ and $ \hat{\bf y}'$
are parallel to two of the $ \hat{\bf n}_k $.
In the following we shall assume for simplicity that this is the case, and in 
particular that $ \hat{\bf x}' = \hat{\bf x} $ and $ \hat{\bf y}' = \hat{\bf y}$.

\begin{figure}[ptb]
\begin{center}
\includegraphics[width=8.5cm]{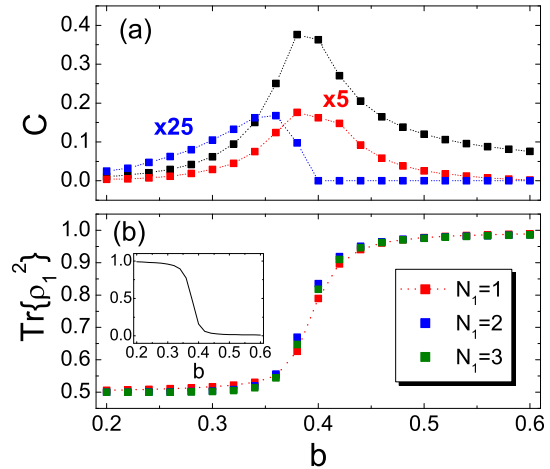}
\end{center}
\caption{(color online) Quantum entanglement in an hexagonal ring of $s=1/2$
spins, model $H^A_X$ with $J^A_{xx}=-1$, in the presence of an external magnetic field 
${\bf b} = b \hat{\bf x}$. 
(a) Pairwise entanglement, quantified by the
concurrence ($C$) between the spins ${\bf s}_k$ and ${\bf s}_l$, with  
$ | k - l | = 1 $(black squares), 2 (red, multiplied by a factor 5), 3 (blue, 
factor 25). 
(b) Block entanglement between subsystems $ \mathcal{S}_1$ and $ \mathcal{S}_2$,
(consisting of $N_1$ and $N_2=N-N_1$ consecutive spins, respectively),
quantified by Tr$ (\rho_1^2) $, with $ \rho_1$ the reduced density matrix of 
$\mathcal{S}_1$. 
Inset: Residual tangle of the single spin. }
\label{fig04a}
\end{figure}
As suggested by the high degree of degeneracy of its energy spectrum (see below),
the collinear Ising models $ H_{X} $ and $ H_{Y} $ are also invariant under a number 
of additional 
transformations, that apply to the orbital or spin degrees of freedom separately.
In the case of the hexagon (Fig. \ref{fig01}), for example, the symmetry 
operations of $H_{\xi = X,Y} $ include all the elements of the ${\bf D}_{6h}$ group, 
where each transformation is applied to the orbital part only: 
6 different $\hat{C}_2$ rotations $ \exp ( -i {\bf L} \cdot \hat{\bf n}_k \pi / \hbar ) $,
with $ {\bf L} = \sum_{i=1}^N {\bf l}_i $ and the 
$6$ rotations axes $ \hat{\bf n}_k $ defined as above; 
reflection about the 
polygon plane $xy$; 
$ \hat{C}_n $ rotation 
$ \exp ( -i 2\pi L_z / N \hbar ) $
around the vertical axis $z$, whih can be thought as the analogue of 
translational invariance for a system with periodic boundary conditions.  
If we assume for 
simplicity that the electrons occupy spherically symmetric orbitals centered
at the positions $ {\bf r}_i $, the transformations of ${\bf D}_{6h}$ simply 
result in permutations of the spin indices. The rotation of 
an angle $ \pi / 3 $ around the $z$ axis, e.g., induces the following 
transformations: 
$ ( s_{k,x} , s_{k,y} , s_{k,z} ) \longrightarrow ( s_{k+1,x} , s_{k+1,y}, s_{k+1,z} ) $.  
In general, one can show that the $ H_{XY} $ Hamiltonian of an $N$-spin regular 
polygon 
is invariant under transformations belonging 
to the 
$ {\bf D}_{nh}$ group (with $n=N$). 
The models $A$ and $B$, instead, are only invariant under reflection of the orbital 
degrees of freedom about the $xy$ plane. 

As far as spin transformations are concerned, the collinear 
$XY$ model belongs to the ${\bf D}_{2h}$ group, with the $C_2$ 
axes that coincide with the $x'$ and $y'$ axes. 
The $\hat{C}_2$ rotations thus correspond to 
$ \exp ( -i S_\alpha \pi / \hbar ) $,
with $ {\bf S} = \sum_{i=1}^N {\bf s}_i $ and 
$ \alpha = x',y',z $.  
The collinear Ising model $H_X$ ($H_Y$) is additionally invariant with respect to
continuous spin rotations around the $x'$ ($y'$) axis.
The Hamiltonians $H_A$ and $H_B$, instead, are only invariant
under reflection about the $xy$ plane, corresponding to the transformation:
$ ( s_{k,x} , s_{k,y} , s_{k,z} ) \longrightarrow ( -s_{k,x} , -s_{k,y}, s_{k,z} ) $.
We finally note that the Hamiltonians $H^A_\xi$ and $H_{\xi'} $ ($ \xi , \xi' = X,Y$) are
unitarily equivalent: $ H_\xi^A = \mathcal{U}_{\xi\xi'} H_{\xi'} \mathcal{U}^{-1}_{\xi\xi'} $. 
In the case $ \hat{\bf x}' = \hat{\bf x} $, for example, 
$ \mathcal{U}_{XX} = \otimes_{k=1}^N \exp ( -i s_{k,z} \phi_k / \hbar ) $.
Therefore, one can associate the symmetry operation
$ \mathcal{U}_{\xi\xi'} C $ of $ H_{\xi'} $ to any symmetry operation of
$ C $ of $ H_{\xi}^A $ (and vice versa). 
As shown below, the collinear and noncollinear models become significantly different
(i.e. no longer unitarily equivalent) in the presence of an applied magnetic field.

\section{Results}

\subsection{Energy spectra and trial wavefunctions}

The results presented below are based on the direct diagonalization of the 
Hamiltonians $H^A$ and $H^B$ for rings of $s=1/2$ spins. 
The energy spectra of the noncollinear 
Ising models (Fig. \ref{fig02}) are independent on whether
$ J_{xx} \neq 0 = J_{yy} $ ($ H^A = H^A_X $)
or
$ J_{yy} \neq 0 = J_{xx} $ ($ H^A = H^A_Y $),
and coincide with those of the collinear Ising model ($ H_{\xi = X , Y } $).
As already mentioned in the previous section, all these models are in fact 
unitarily equivalent. 
The twofold degenerate ground states of $ H^A_\xi $, like any other 
eigenstate $ | \Psi_k^A \rangle $, can be derived from those of $ H_{\xi'} $
by the unitary transformation $\mathcal{U}_{\xi\xi'}$: 
\begin{equation}\label{eqbeta}
| \beta_{k}^\chi ( \phi ) \rangle = \mathcal{U}_{\xi\xi'} 
| \alpha_x^\chi \rangle = \left[ \otimes_{l=1}^N R^l_z (\phi_l +\phi + k\pi) \right] 
| \alpha_x^\chi \rangle ,
\end{equation}
where $ R_{k,z} (\varphi) = \exp (-i s_{k,z} \varphi ) $, $ \phi = \phi (\xi,\xi') $,
and $k=0,1$. 
The expression of $ | \alpha_x^\chi \rangle $ depends on whether  
the coupling has a ferromagnetic ($ \chi = F $) or an antiferromagnetic ($ \chi = AF $) character:
\begin{eqnarray}\label{alfa}
| \alpha_x^F   \rangle = | \uparrow_x \uparrow_x\dots\uparrow_x\uparrow_x \rangle ,
\nonumber\\  
| \alpha_x^{AF} \rangle = | \uparrow_x\downarrow_x\dots\uparrow_x\downarrow_x \rangle ,
\end{eqnarray}
where $ | \uparrow_x \rangle $ and $ | \downarrow_x \rangle $ are the eigenstates of
the single-spin projection along the $x$ direction.
Finally, one can easily verify that $ \phi = 0 $ for $ \xi = \xi' $, whereas
$ \phi = \pm\pi / 2 $ if $ (\xi , \xi' ) $ coincides with $ ( Y , X ) $ or $ ( X , Y ) $,
respectively. 

The spectrum of $H^B_\xi $ is characterized by a lower degree of degeneration with respect 
to that of $H^A_\xi $ (Fig. \ref{fig02}), reflecting the lower symmetry of the 
former Hamiltonian with respect to the latter one. 
In particular, the ground state doublet presents a splitting $ \delta $, 
whose magnitude decreases with the number of spins, as 
reported in the figure caption.
For each spin ring, the energy spectrum is independent on whether 
$ H^B = H^B_X $ or $ H^B = H^B_Y $ and on the sign of the coupling $ J^B $. 
In fact, all these Hamiltonians are unitarily equivalent, being: 
\begin{eqnarray}
H^B_Y\!\! & = &\!\! \mathcal{U}\, H^B_X\, \mathcal{U}^{-1},\ {\rm with\ } \mathcal{U} =
\otimes_{k=1}^N e^{-i\pi s_{k,z}/2} , \\
H^B_X \!\! & = &\!\! \mathcal{U}\, (-H^B_X)\, \mathcal{U}^{-1},\ {\rm with\ } \mathcal{U} =
\otimes_{k=1}^{N/2} e^{-i\pi s_{2k,z}} , 
\end{eqnarray}
where the latter equation also implies that the energy spectrum is symmetric with respect 
to the origin.
In the following
we thus refer, without loss of generality, to the case of $ H^B_X $ with $ J_X^B > 0 $.
In all the considered cases, it was found that the ground state $ H^B_X $ could be 
expressed as a linear combinations of a limited number of symmetry-adapted states:
\begin{eqnarray}\label{psi0B}
| \Psi_0^B \rangle = \left[\otimes_{l=1}^{N} R_{l,z} \left( \phi_l \right) \right]
\sum_{k=0}^{N/2} 
\sum_{\{ {\bf v}_{k} \}} C_{k}^{{\bf v}_{k}} | \Phi^{{\bf v}_{k}}_{k} \rangle ,
\end{eqnarray}
where the total spin projection along $z$ of each component is fixed by $k$ ($M=N/2-2k$).
The components $ | \Phi^{{\bf v}_{k}}_{k} \rangle $, whose coefficients 
$ C_{k}^{{\bf v}_{k}} $ are determined numerically, are given by
\begin{eqnarray}\label{eq01}
| \Phi^{{\bf v}_{k}}_{k} \rangle = 
 (-1)^{\sum_p v^{k}_p } \sum_n \left( \otimes_{q=1}^{2k} \sigma_{v^{k}_q+n,x} \right) 
| \alpha_z^F \rangle ,
\end{eqnarray}
where the $2k$ elements $ 1 \le v^k_p \le N $ of the vector ${\bf v}_k$ specify which spins are flipped 
with respect to the reference configuration $ | \alpha_z^F \rangle = 
| \uparrow_z \uparrow_z\dots\uparrow_z\uparrow_z \rangle $ .
In Eq. \ref{eq01}, different vectors ${\bf v}_k$ correspond to components that 
cannot be transformed one into another by rotating the spin ring of an angle 
$ \phi_l $ around the $z$ axis. 
For example, the components of
$ | \Phi^{{\bf v}_{1}}_{1} \rangle $ and $ | \Phi^{{\bf v}_{2}}_{1} \rangle $ can be represented by all the states where the only two down spins are nearest neighbours
or next nearest neighbours, respectively.
In the case $ J_X < 0 $, the state $ | \alpha_z^F \rangle $ in Eq. \ref{eq01}
is replaced by its antiferromagnetic counterpart $ | \alpha_z^{AF} \rangle $.
Besides, additional relations are found between the coefficients of 
$ | \Phi^{{\bf v}_{k}}_{k} \rangle $ 
and 
$ | \Phi^{\tilde{\bf v}_{k}}_{N/2-k} \rangle $, that are the spin-flipped versions 
of one another, depending on the model, direction and character (ferromagnetic or
antiferromagnetic) of the coupling (Table \ref{tableI}).
The use of the expression \ref{psi0B} as a trial wavefunction for the ground state
allows to reduce drastically the dimension of the Hamiltonian to be diagonalized,
for example from 256 to 12 for $N=8$ spins $s=1/2$, or from 1024 to 15 for $N=10$. 

\begin{table}
\begin{tabular}{|c|cccc|}
\hline
$ C_{k}^{{\bf v}_{k}} / C_{k}^{\tilde{\bf v}_{N/2-k}}$ & $ J_{xx} > 0 $ & $ J_{xx} < 0 $ & $ J_{yy} > 0 $ & $ J_{yy} < 0 $ \\
\hline
model $A$ & -1 &  1 &  1 & -1 \\
model $B$ &  1 & -1 & -1 &  1 \\
\hline
\end{tabular}
\caption{\label{tableI} Ratio between the coefficients $ C_{k}^{{\bf v}_{k}} $ and
$ C_{k}^{\tilde{\bf v}_{N/2-k}} $ in the ground state of the noncollinear 
Ising models $B$. The components $ | \Phi^{{\bf v}_{k}}_{k} \rangle $ and 
$ | \Phi^{\tilde{\bf v}_{k}}_{N/2-k} \rangle $  in Eq. \ref{psi0B} are the 
spin-flipped versions of one another.}
\end{table}

\subsection{Ground state entanglement without field}

Unlike the standard Ising model, the noncollinear one ($ H^B_X $) presents remarkable entanglement properties in the absence of an external magnetic field.
In fact, we find that the ground state of $H^B$ approximately corresponds to a symmetric 
combination of 
the two degenerate ground states of $H^A$:
\begin{eqnarray}\label{eqpsiGHZ}
| \widetilde{\Psi}^\xi (\phi) \rangle = \frac{1}{\sqrt{2}} \left[ | \beta_{0}^\xi (\phi)\rangle +
                                              | \beta_{1}^\xi (\phi)\rangle \right] ,  
\end{eqnarray}
where $ \phi = \pi / 2 $ and $ \xi = F $ ($ \xi = AF $) for $ J^B_{xx} $ negative 
(positive). 
As reported in Fig. \ref{fig03} (filled blue squares), the squared modulus of such overlap, $ p = | \langle \Psi_0^B | \widetilde{\Psi}^\xi \rangle |^2 $, increases with the spin number $N$ and approaches 
1 already for $N=10$.  
As a consequence, the ground state of the noncollinear Ising model $B$ essentially 
corresponds to a linear superpositions of macroscopically different components, where the state of each spin in $ | \beta_{0}^\xi (\phi) \rangle $ is orthogonal to the state of the same spin in 
$ | \beta_{1}^\xi (\phi) \rangle $. 
In analogy with the magnetization and N\'eel vectors, whose coherent tunneling is expected to take 
place in the ground state of  
molecular nanomagnets with ferromagnetic and antiferromagnetic Heisenberg interaction, respectively \cite{gatteschi}, 
we introduce here the vectors $ {\bf n}_F $ and 
$ {\bf n}_{AF} $, defined as:
\begin{eqnarray}
{\bf n}_F ( \phi )\!\!\! & = &\!\!\! \frac{1}{Ns} \sum_{k=1}^N \left\{ 
R^{-1}_{k,z} (\phi_k +\phi)\, {\bf s}_{k}\, R_{k,z} (\phi_k +\phi) 
\right\} ,\\ 
{\bf n}_{AF} ( \phi )\!\!\! & = & \!\!\! \sum_{k=1}^N \!\frac{(-1)^k}{Ns}\! \!\left\{\! 
R^{-1}_{k,z} (\phi_k \!+\!\phi)\, {\bf s}_{k}\, R_{k,z} (\phi_k \!+\!\phi) 
\!\right\}\! .
\end{eqnarray}
These vectors have maximum modulus and opposite orientations in the case of the 
two macroscopically different components in Eq. \ref{eqpsiGHZ}:
$ | \langle \beta_0^\xi (\phi) | {\bf n}_\xi ( \phi ) | \beta_0^\xi (\phi) \rangle |
= 1 $ and
$ \langle \beta_0^\xi (\phi) | {\bf n}_\xi ( \phi ) | \beta_0^\xi (\phi) \rangle
=
- \langle \beta_1^\xi (\phi) | {\bf n}_\xi ( \phi ) | \beta_1^\xi (\phi) \rangle $,
with $ \xi = F, AF $.
We can thus summarize the result reported in Fig. \ref{fig03} by saying that the ground
state of the noncollinear Ising models $ H^B_X $ presents coherent tunneling of the 
vectors $ {\bf n}_\xi ( \pi / 2 ) $.  

Similar results can be found in the case of odd spin numbers. 
Here, the overall Hilbert space can be divided into two uncoupled subspaces, including either
the states with $ M = - N/2 + 2(k-1) $, or those with $ M = + N/2 - 2(k-1) $ (being 
$k=1,2,\dots , N/2 $). 
This results in a twofold degeneracy of all eigenvalues. 
Besides, the ferromagnetic and antiferromagnetic cases are no longer equivalent,
and an additional degeneracy of the ground state is induced by spin frustration in the 
latter case. 
In Fig. \ref{fig03}, we report the squared modulus of the overlap $ \langle \Psi_0^B | \widetilde{\Psi}^F \rangle $ for odd $N$ (stars), where $ | \Psi_0^B \rangle $ is 
obtained by diagonalizing $H^B_X$ within each of the above mentioned subspaces.  
The trend as a function of $N$ resembles that obtained for even spin numbers. 
The possibility of approximating the ground state of $H^B$ with the state 
$ | \widetilde{\Psi}^\xi (\phi) \rangle $ applies also to spins $ s > 1/2 $.
In these cases, the single-spin states $ | \uparrow_x\rangle $ and 
$ | \downarrow_x\rangle $ that enter the definition of $ |\alpha_x^\chi\rangle $
(Eq. \ref{alfa}) correspond to $ | m_x = +s \rangle $ and 
$ | m_x = -s \rangle $, respectively. 
For example, from analogous calculations performed on $s=1$ spins (not reported 
here) the overlap between the ground state of $H^B$ and 
$ | \widetilde{\Psi}^\xi (\phi) \rangle $ is larger than for rings of $1/2$ spins
with equal $N$.

\begin{figure}[ptb]
\begin{center}
\includegraphics[width=8.5cm]{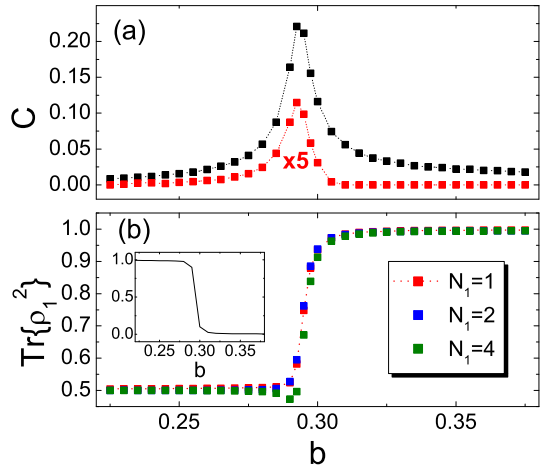}
\end{center}
\caption{(color online) Quantum entanglement in an octagonal ring of $s=1/2$
spins, model $H^A_X$ with $J^A_{xx}=-1$, in the presence of an external magnetic field 
${\bf b} = b \hat{\bf x}$. 
(a) Pairwise entanglement, quantified by the
concurrence ($C$) between the spins ${\bf s}_k$ and ${\bf s}_l$, with  
$ | k - l | = 1 $ (black squares), 2 (red, multiplied by a factor 5). 
(b) Block entanglement between subsystems $ \mathcal{S}_1$ and $ \mathcal{S}_2$,
(consisting of $N_1$ and $N_2=N-N_1$ consecutive spins, respectively),
quantified by Tr$ (\rho_1^2) $, with $ \rho_1$ the reduced density matrix of 
$\mathcal{S}_1$. 
Inset: Residual tangle of the single spin. }
\label{fig04b}
\end{figure}

From the point of view of quantum correlations, the states 
$ | \widetilde{\Psi}^\xi \rangle $ are equivalent to the Greenberger-Horne-Zeilinger
(GHZ) states, characterized by a genuine multipartite entanglement and by vanishing
pairwise entanglement.
Pairwise entanglement between nearest neighbouring spins is however present in the 
ground state of $ H^B_X $.
In fact, the concurrence between nearest neighbours has finite values 
(red squares in Fig. \ref{fig03}), that
decrease for increasing $N$, and tend to zero as the ground state tends to the 
GHZ-like state 
$ | \widetilde{\Psi}^\xi (\phi) \rangle $. 
The concurrence between pairs of spins that aren't nearest neighbours (not shown)
is zero in all the considered cases. 

\subsection{Magnetic field induced entanglement}

The magnetic field can be used as a control parameter in order to tune the 
quantum correlations within the ring. 
In the case of model $A$, the two degenerate and separable ground states 
$ | \beta_{k}^\chi ( \phi ) \rangle $ can be coupled by applying an homogeneous
in-plane magnetic field $ H_b^A = b \sum_{k=1}^N s_{k,x} $. 
As a result, the degeneracy is removed, and  
$ | \Psi^A_0 \rangle $ tends to a linear superposition of macroscopically 
distinct states. The overlap between the ground state and 
$ | \widetilde{\Psi}^\xi (\phi) \rangle $ (Eq. \ref{eqpsiGHZ})
as a function of the magnetic field is reported in Fig. \ref{fig04a} 
($N=6$) and Fig. \ref{fig04b} ($N=8$).
The in-plane magnetic field reduces the symmetry of the system and breaks
the equivalence between the noncollinear Ising models $H_\xi^A$ and their
collinear counterparts, and that between the ferromagnetic and 
antiferromagnetic cases. 
One of the consequences of the symmetry reduction is the removal of the 
degeneracy in the ground state. In fact, the energy splitting $ \delta = E_1 - E_0 $ 
between ground and first 
excited states, increases with $ b$ (not shown). 
The field however also mixes the subspace
spanned by $ | \beta_0^\chi ( \phi ) \rangle $ and 
$ | \beta_1^\chi ( \phi ) \rangle $ with additional components, thus reducing the
overlap between $ | \Psi^A_0 \rangle $ and $ | \widetilde{\Psi}^\xi (\phi) \rangle $
(see Table \ref{tableII}).
\begin{figure}[ptb]
\begin{center}
\includegraphics[width=8.5cm]{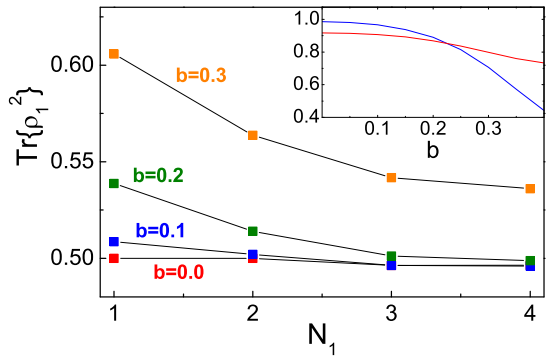}
\end{center}
\caption{(color online) (color online) Entanglement between two subsystems
of consecutive spins ($\mathcal{S}_1$ and $\mathcal{S}_2$) quantified by 
$ {\rm Tr} (\rho_1^2) $, being $ \rho_1$ the reduced density matrix of 
$\mathcal{S}_1$. The trace is computed for the ground state of $H^B+H^B_b$,
and is displayed as a function of the number of spins of $\mathcal{S}_1$, 
$N_1=N-N_2$, in the case $N=8$ and for different values of the magnetic 
field. Inset: Residual tangle (blue curve) and concurrence $C$ (plotted 
as $1-C$, red) as a function of the magnetic field $b$.}
\label{fig05}
\end{figure}
All this is reflected in the entanglement properties of the ground state.
For low values of the field, the concurrence between all pairs of spins is 
negligible [Figs. \ref{fig04a}(a) and \ref{fig04b}(a)], whereas any blocks of
$N_1$ consecutive spins is entangled with the complementary block of 
$ N_2 = N - N_1 $ (panels (b)). Moreover, the value of Tr$(\rho_1^2)$ is 
close to 1/2, independently on the partition ($N_1$). 
Finally, the residual tangle, that quantifies the genuine multipartite 
entanglement and is defined as 
$ 4 $det$ ( \rho_k ) - \sum_{i \neq k} C_{ik}^2 $ (with $ \rho_k $ the reduced
density matrix of the $k$-th spin $C_{ik}$ the concurrence between spins 
$i$ and $k$),
is maximized (insets).
All these features are consistent 
with a GHZ-like form of the ground state. For high values of the field, 
the ground 
state tends to a factorizable form, as shown by the simultaneous suppression 
of pairwise, block and multipartite entanglement. 
In the intermediate region ($ b \simeq 0.4 $ for $N=6$ and $ b \simeq 0.3 $ 
for $N=8$), the abrupt reduction of the residual tangle and of the block 
entanglement is accompanied by peaks in the values of the concurrence, not 
only between nearest neighbours.
 
We note in passing that the noncollinear Ising $H^A_X$, combined with an 
homogeneous magnetic field, results in an Hamiltonian and in ground state 
entanglement properties that are equivalent to those  
of a collinear Ising interaction in the presence of an inhomogeneous 
magnetic field, with radial field orientation at each spin site 
($ {\bf b}_k \parallel {\bf r}_k $). 
While the latter geometry might produce remarkable entanglement properties in 
mesoscopic (pseudo)spin systems \cite{PhysRevLett.100.100502},
the former one seems much more suitable for producing analogous effects 
in nanometer-sized objects, such 
as molecular nanomagnets.

In the case of model $B$, we consider a magnetic field applied along the $z$ 
direction, giving rise to an additional term in the Hamiltonian:
$ H_b^B = b \sum_{k=1}^N s_{k,z} $.
Such field preserves the equivalence between the ferromagnetic ($ J^B_{xx}<0 $)
and antiferromagnetic ($ J^B_{xx}>0 $) models, as well as that between 
the $ J^B_{xx} = 0 $ and $ J^B_{yy} = 0 $ cases.
We find that, also in the presence of the field, the ground state of the 
noncollinear Ising model $B$ is well approximated by a linear
superposition of two macroscopically different states:
\begin{eqnarray}\label{eqpsiGGHZ}
| \widetilde{\Psi}^\xi_t (\theta , \phi) \rangle = 
\frac{1}{\sqrt{C}} \left[ 
| \gamma_{0}^\xi (\theta , \phi)\rangle +
| \gamma_{1}^\xi (\theta , \phi)\rangle \right] ,
\end{eqnarray}
where
\begin{equation}
| \gamma_k^\xi ( \theta , \phi ) \rangle 
= \left[ \otimes_{l=1}^N R_{l,z} (\phi_l +\phi+ k\pi) \right] 
| \alpha_t^\xi (\theta ) \rangle .
\end{equation}
Here $ \xi = F $ or $ \xi = AF $, depending on whether the coupling has a 
ferromagnetic or an antiferromagnetic character:
\begin{eqnarray}
| \alpha_t^F   (\theta ) \rangle & = & \left[ \otimes_{q=1}^N R_{q,y} ( \theta ) \right]
| \uparrow_z\uparrow_z\dots\uparrow_z \rangle ,
\nonumber\\  
| \alpha_t^{AF}(\theta ) \rangle & = & \left\{ \otimes_{q=1}^N R^q_y [ (-1)^q\theta ] \right\} | \uparrow_z\uparrow_z\dots\uparrow_z \rangle .
\end{eqnarray}
Unlike $ | \beta_0^\xi ( \phi ) \rangle $ and $ | \beta_1^\xi ( \phi ) \rangle $
(Eq. \ref{eqbeta}), the states
$ | \gamma_0^\xi ( \theta , \phi ) \rangle $ and
$ | \gamma_1^\xi ( \theta , \phi ) \rangle $ 
are not mutually orthogonal, unless
$ \theta = \pi /2 $, so that
$ | \alpha_t^F (\theta ) \rangle = | \alpha_x^F (\theta ) 
\rangle $ and $ | \gamma_k^\xi ( \theta , \phi ) \rangle = | \beta^\xi_k ( \phi ) \rangle $.
In general,
$ | \langle \gamma_0^\xi ( \theta , \phi ) | \gamma_1^\xi ( \theta , \phi ) \rangle 
= | \cos\theta |^N $; the normalization constant in Eq. \ref{eqpsiGGHZ} is 
thus $ C = 2(1+| \cos\theta |^N) $.
As detailed in Table \ref{tableII}, for increasing values of the field, the overlap 
between $ | \Psi^B_0 \rangle $ and $ | \widetilde{\Psi}^\xi_t (\theta , \phi) \rangle $
decreases, while the value $ \theta_M $ of the angle $ \theta $ that maximizes such overlap
decreases. As $ \theta_M $ passes from $ \pi /2 $ (like in the case $b=0$)
to lower values, the entanglement properties of the ground state deviate from those
of a GHZ state (Fig. \ref{fig05}). 
In particular, the entanglement between a block of $ N_1 $ consecutive spins and
the remaining $ N_2 = N - N_1 $ ones, quantified by Tr$\{ \rho_1^2 \}$, 
becomes an increasing function of the $N_1$, whereas for $b=0$ it is practically 
independent on $N_1$, as for a GHZ state. 
The pairwise entanglement between nearest neighbouring spins, quantified by the 
concurrence, increases (red curve, figure inset). 
The multipartite entanglement, as quantified by the residual tangle, decreases for 
increasing $b$ (blue curve).
Altogether, the perpendicular magnetic field thus induces a transition from a 
predominantly multipartite entangled ground state to one with large pairwise quantum 
correlations.
\begin{table}
\begin{tabular}{|c|c|ccccc|}
\hline
\hline
\multicolumn{2}{|c|}{ $ b\, ( {\bf b} \parallel \hat{\bf x}) $ } 
& $ 0.2\ $ & $ 0.225\ $ & $ 0.25 \ $ & $ 0.275\ $ & $ 0.3\ $ \\
\hline
model & $p$ & 0.956 &  0.947 &  0.936 & 0.913 & 0.0595 \\
$A$ & $ \ \theta_M / \pi \ $  &  0.5 & 0.5 & 0.5 &  0.5 & 0.5 \\
\hline
\hline
\multicolumn{2}{|c|}{ $ b\, ({\bf b} \parallel \hat{\bf z}) $ } 
& $ 0\ $ & $ 0.1\ $ & $ 0.2\ $ & $ 0.3\ $ & $ 0.4\ $ \\
\hline
model & $p$ & 0.985 &  0.980 &  0.960 & 0.909 & 0.862 \\
$B$ & $ \ \theta_M / \pi \ $  &  0.250 & 0.229 & 0.206 &  0.173 & 0.136 \\
\hline
\hline
\end{tabular}
\caption{\label{tableII} Square modulus $p$ of the overlap between the trial 
wavefunction 
$ | \widetilde{\Psi}^\xi_t (\theta , \phi) \rangle $
and ground state of $H^A+H_b^A$ (upper lines) or $H^B+H_b^B$ 
(lower lines), with $N=8$. 
The angle $\theta_M$ is the value of $\theta$ that maximizes $p$.
In the case of the models $A$ and $B$, the magnetic field
is oriented along the $x$ and $z$ directions, respectively. 
The above results refer to $H^A_X$, with $ J_{xx}^A = -1 $; as to
the model $B$, no difference emerges between 
$ H^B_X + H^B_b $ and $ H^B_Y  + H^B_b $, 
nor between $ J_{xx}^B = +1 $ and $ J_{xx}^B = -1 $.}
\end{table}

\section{Conclusions}

In conclusion, we have investigated spin rings that are coupled by noncollinear Ising 
interactions, whose anisotropy reflects the point-group symmetry of the system. 
The ground states of these Hamiltonians exhibit remarkable entanglement properties.
In particular, in the case where the preferential directions for each spin are 
determined by the direction of the spin-spin bond (model $B$), the system ground
state $ | \Psi_0^B \rangle $ is characterized by a large multipartite entanglement
and by a low degree of pairwise entanglement. In fact, the overlap between 
$ | \Psi_0^B \rangle $ and a GHZ-like state increases with the number of spins $N$,
and approaches 1 already for $N=10$. A vertical magnetic field can be 
used to substantially modify such picture, enhancing the pairwise entanglement 
between nearest neighbouring spins at the expense of multipartite entanglement. 
In the case where the preferential directions for each spin are determined solely 
by its position within the ring (irrespective of the bond direction) - model $A$ -
the noncollinear Ising Hamiltonian is unitarily equivalent to the standard Ising 
model, and thus the degenerate ground state doublet is spanned by two factorizable
states. However, the application of an moderate (with respect to $J$) in-plane field 
splits such doublet and induces multipartite entanglement in the ground state
$ | \Psi_0^A \rangle $.
For increasing values of the field, $ | \Psi_0^A \rangle $ undergoes a sharp 
transition towards a separable ferromagnetic ground state, accompanied a peak in 
the pairwise entanglement, not only between nearest neighbouring spins.
While these results have been obtained for $s=1/2$ spins, analogous behaviours 
emerge from preliminary calculations performed with higher spin values.

\end{document}